\documentclass[10pt,conference]{IEEEtran}
\IEEEoverridecommandlockouts

\usepackage{caption}
\usepackage[normalem]{ulem}
\usepackage{cite}
\usepackage{algorithm, algorithmic, lipsum}
\usepackage{amsmath,amssymb, amsfonts, dsfont, braket}
\usepackage{mathtools, braket, lipsum, bibentry}
\usepackage{graphicx, multirow, tabularx, hyperref}
\usepackage{textcomp}
\usepackage[usenames,dvipsnames]{xcolor}
\usepackage[utf8]{inputenc} 
\usepackage[T1]{fontenc}    
\usepackage{booktabs}       
\usepackage{nicefrac}       
\usepackage{microtype}      
\usepackage{float}
\usepackage{enumitem, lipsum}
\usepackage[caption=false,font=footnotesize]{subfig}
\usepackage{xfrac} 

\def\BibTeX{{\rm B\kern-.05em{\sc i\kern-.025em b}\kern-.08em
    T\kern-.1667em\lower.7ex\hbox{E}\kern-.125emX}}
\newcommand{\xrm}{\textrm{x}}

\hypersetup{colorlinks=true, linkcolor=black, urlcolor=blue}

\graphicspath{
{figures/}
{/home/samudra/Desktop/QuantumDissertation/source_materials/archive_QuantumReliability/archived_figures/}
{/home/samudra/Desktop/QuantumDissertation/source_materials/QuantumReliability/QuantumReliability_acm2023_github/figures/}
}
 
\begin{document}
\title{Reliable Devices Yield Stable Quantum Computations
\thanks{The manuscript is authored by UT-Battelle, LLC under Contract No.~DE-AC05-00OR22725 with the U.S. Department of Energy. The U.S.~Government retains for itself, and others acting on its behalf, a paid-up nonexclusive, irrevocable worldwide license in said article to reproduce, prepare derivative works, distribute copies to the public, and perform publicly and display publicly, by or on behalf of the Government. The Department of Energy will provide public access to these results of federally sponsored research in accordance with the DOE Public Access Plan: https://www.energy.gov/doe-public-access-plan.}
}

\author{
\IEEEauthorblockN{Samudra~Dasgupta$^{1, 2^*}$, and Travis S.~Humble$^{1,2^\dagger}$\\}
\IEEEauthorblockA{
\textit{$^1$Quantum Science Center, Oak Ridge National Laboratory, Oak Ridge, Tennessee, USA}\\
\textit{$^2$Bredesen Center, University of Tennessee, Knoxville, Tennessee, USA}\\
\textit{$^*$sdasgup3@tennessee.edu, ORCID: 0000-0002-7831-745X}\\
\textit{$^\dagger$humblets@ornl.gov, ORCID: 0000-0002-9449-0498}\\
}}

\maketitle

\begin{abstract}
Stable quantum computation requires noisy results to remain bounded even in the presence of noise fluctuations. Yet 
non-stationary noise processes lead to drift in the varying characteristics of a quantum device that can greatly influence the circuit outcomes.
Here we address how temporal and spatial variations in noise relate device reliability to quantum computing stability. First, our approach quantifies the differences in statistical distributions of characterization metrics collected at different times and locations using Hellinger distance. We then validate an analytical bound that relates this distance directly to the stability of a computed expectation value. Our demonstration uses numerical simulations with models informed by the transmon device from IBM called washington. We find that the  stability metric is consistently bounded from above by the corresponding Hellinger distance, which can be cast as a specified tolerance level. These results underscore the significance of reliable quantum computing devices and the impact for stable quantum computation. 
\end{abstract}


\section{Introduction}
Quantum devices are subject to non-stationary noise sources, e.g. non-uniform spontaneous decay, energy loss, cross-talk, sensitivity to imprecise control pulses, and fluctuations in thermodynamic controls, all of which affect the quality of the quantum register implementation. The field of quantum noise characterization focuses on measuring and tracking noise metrics (such as CNOT gate error) at various points in time. These characterizations inform calibration techniques for hardware engineers as well as error mitigation methods for programmers.
However, quantum devices also exhibit temporal variations in their noise sources, which underlies the need for frequent calibration and adjustment of device metrics. Non-stationary noise processes can also stymie attempts at characterization as the underlying noise models must adapt to new and often unpredictable behaviors. How can we monitor changes in the noise itself to better inform these efforts?
\par 
Here, we address the concern that non-stationary noise process pose to reliable quantum computation. Device reliability is presented as a measure of the statistical similarity of the underlying device metrics, such as gate fidelities and coherence times. This measure captures the similarity between device metrics considering both spatial and time-varying noise processes. We then recall how device reliability bounds the stability of expectation values dervied from noisy quantum computation. Moreover, we validate this bound on stablility using numerical simulations of a circuit modeled by a multi-dimensional correlated noise distribution.
\section{Theory}
\subsection{Stability}
Stability in quantum computing refers to the boundedness of the output of a quantum circuit in the presence of noise fluctuations \cite{dasgupta2022assessing}. In this study, we focus on the mean value of a quantum observable $O$ as a representative of program output. Let $\xrm$ represent the parameter characterizing the noisy realization of a quantum circuit $\mathcal{C}$. The mean value of $O$ obtained from repeated executions on the noisy circuit is denoted as $\braket{O}_\xrm$. Considering the time-varying nature of device noise, we introduce $f(\xrm; t)$ as the probability distribution function of the quantum noise parameter $\xrm$. We define $\braket{O}_t$ as the average value of $\braket{O}_\xrm$ with respect to $f(\xrm; t)$, the probability distribution function for the noise parameter. 
\begin{equation}
\braket{O}_t = \int\limits \braket{O_{ \xrm}}  f(\xrm; t) d\xrm
\label{eq:O_td_def}
\end{equation}
The stability of the quantum observable between two time points, $t_1$ and $t_2$, is then quantified by the absolute difference in the mean values of $\braket{O}$ obtained at those times, defined as
\begin{equation}
s(t_1, t_2) = | \braket{O}_{t_1} - \braket{O}_{t_2} |
\label{eq:stability}
\end{equation}
\subsection{Reliability}
We next quantify device reliability by comparing the statistical distributions of various characterization metrics collected at different times and register locations. When these metrics exhibit statistical similarity, the device behavior is considered to be reliable. The statistical distance between distributions is calculated using the Hellinger distance, which offers ease of calculation and interpretation. 
\begin{equation}
H( f(\xrm ; t_1), f(\xrm ; t_2) ) = \sqrt{1-\int\limits_{\xrm} \sqrt{f(\xrm ; t_1) f(\xrm ; t_2)} d\xrm}
\label{eq:hellinger_unmodified}
\end{equation} 
The Hellinger distance above quantifies the statistical similarity of a device at different times such that when the distance is small, the device behaves statistically similar at both times. This is expected when the underlying noise process is stochastic. However, larger values of the distance imply that noise processes within the device are non-stationary processes that lead to noticeable changes in device properties. The timescales on which such statistically significant changes are measured represent an important metric for evaluating the reliability of the device relative to a desired tolerance.
\subsection{Stability Bounds}
We now establish an analytical and intuitive connection between output stability and device reliability. Specifically, we show how device reliability constrains the outcomes from a quantum program executed on a NISQ device by examining the role of fluctuations in device metrics. 
\par 
Let $s_\textrm{tol}$ denote a specified tolerance on the stability metric introduced earlier. Additionally, let the reliability of the quantum device between times $t_1$ and $t_2$ is quantified by the Hellinger distance $H_X$, as discussed previously. We determine the maximum bound $H_X^{\textrm{max}}(t_1, t_2)$ on the Hellinger distance constrained by $s(t_1, t_2) < s_\textrm{tol}$. 
\par 
We begin by noting the bound on the stability satisfies
\begin{equation}
s^2(t_1, t_2) \leq \left( \int \left| \braket{O_{ \xrm}}  \{ f(\xrm; t_1)\textrm{dx} - f(\xrm; t_2)\} \right| \textrm{dx} \right)^2
\label{eq:abs_integrand}
\end{equation}
where the inequality stems from the absolute value on the integrand. Per Holder's inequality, if $m, n \in [1, \infty]$ and $\sfrac{1}{m}+\sfrac{1}{n}=1$ then
\[ \int \left| f(x) g(x) \right|dx  \leq  \left( \int |f(x)|^m dx\right)^{1/m} \left( \int |g(x)|^n dx\right)^{1/n}\]
Thus, the right-hand side of Eq.~(\ref{eq:abs_integrand}) becomes
\begin{equation*}
\begin{aligned}
&\left( \int \left| \braket{O_{ \xrm}}  \{ f(\xrm; t_1)\textrm{dx} - f(\xrm; t_2)\} \right| \textrm{dx} \right)^2 \\
&\leq \left( \int  | \braket{O_{ \xrm}} |^m \textrm{dx}\right)^{2/m} \left( \int | f(\xrm; t_1) - f(\xrm; t_2) |^n \textrm{dx}\right)^{2/n}
\end{aligned}
\end{equation*}
Choose $m =\infty, n=1$ and define
$c= \underset{\xrm}{\textrm{sup}} |\braket{O_{ \xrm}}|$.
This circuit-specific constant satisfies
\begin{equation*}
\begin{aligned}
&\lim\limits_{m \rightarrow \infty} \left( \int |\braket{O_{ \xrm}}|^m\textrm{dx}\right)^{1/m}\leq \lim\limits_{m \rightarrow \infty} \left( \int c^m\textrm{dx}\right)^{1/m}= c\\
\label{eq:c_factor}
\end{aligned}
\end{equation*}
Thus, we may then reduce Eq.~(\ref{eq:abs_integrand}) as
\begin{equation*}
\begin{aligned}
s(t_1, t_2)^2 \leq  & \lim\limits_{m\rightarrow \infty, n = 1} 
\left( \int  | \braket{O_{ \xrm}} |^m \textrm{dx}\right)^{2/m}\\
&\left( \int | \{ f(\xrm; t_1)\textrm{dx} - f(\xrm; t_2)\} |^n \textrm{dx}\right)^{2/n} \\[2ex] 
\leq c^2& \left( \int \left| \sqrt{f(\xrm; t_1)} - \sqrt{f(\xrm; t_2)} \right| \right. \\
&\left. \left(  \sqrt{f(\xrm; t_1)} + \sqrt{f(\xrm; t_2)} \right) \textrm{dx} \right)^2\\[2ex] 
\leq c^2 &\int \left( \sqrt{f(\xrm; t_1)} - \sqrt{f(\xrm; t_2)} \right)^2 \textrm{dx} \\
&\int\left( \sqrt{f(\xrm; t_1)} + \sqrt{f(\xrm; t_2)} \right)^2 \textrm{dx}
\end{aligned}
\end{equation*}
Using Holder's inequality with $m=n=2$, this yields $s(t_1, t_2) = 2cH\sqrt{2-H^2}$ which can be re-arranged to yield the maximum
\begin{eqnarray}
H_{\textrm{max}}(t_1, t_2) = \sqrt{1-\sqrt{1-\phi}}
\label{eq:bound}
\end{eqnarray}
with $\phi = s_\textrm{tol}^2 / (4c^2)$. This sets an upper limit $H_{\textrm{max}}$ on the Hellinger distance to ensure the desired stability criterion $s_{\textrm{tol}}$ is met.
\section{Validation}
\subsection{Experimental data}
We utilized data obtained from the transmon device called washington, a 127-qubit register with heavy hexagonal connectivity developed by IBM. The publicly available characterization data for the washington device was used to create a dataset comprising specific device metrics (refer to Table 1). This dataset was constructed from a subset of the device characterization data spanning a 16-month period from January 1, 2022, to April 30, 2023. The Qiskit software library \cite{ibm_quantum_experience_website} was employed to access the collected characterization data online.
\par These metrics correspond to the minimum requirements for the physical implementation of quantum computing \cite{divincenzo2000physical}, which fall into one of the five classes: SPAM (state preparation and measurement) fidelity, single-qubit gate fidelity, two-qubit entangling gate fidelity, duty cycle (gate length to coherence time ratio), and addressability (ability to measure a register element without interference from other qubits). Specifically, these 16 metrics capture the noise processes of the five-qubits employed in the test circuit illustrated in Fig.~\ref{fig:bv_qiskit_ckt}. Our simulation of the test circuit (described in the next section) relies on data pertaining to these 16 metrics, which enables us to estimate the time-varying joint distribution of circuit noise. Utilizing this estimated distribution, Monte Carlo sampling is performed to simulate the test circuit and validate the theory presented earlier.
\begin{table}[htp]
\centering
\begin{tabular}{|p{1.3cm}|p{4.0cm}|p{1.0cm}|}
    \hline 
\textit{Parameter} & \textit{Description} &\textit{Model} \\ \hline
$\xrm_{0}$ & SPAM  fidelity, register  0 & ABC  \\ \hline
$\xrm_{1}$ & SPAM  fidelity, register  1 & ABC  \\ \hline
$\xrm_{2}$ & SPAM  fidelity, register  2 & ABC  \\ \hline
$\xrm_{3}$ & SPAM  fidelity, register  3 & ABC  \\ \hline
$\xrm_{4}$ & CNOT fidelity, control 0, target 1 & DP$\otimes$DP  \\ \hline
$\xrm_{5}$ & CNOT fidelity, control 2, target 1  & DP$\otimes$DP \\ \hline
$\xrm_{6}$ & $T_2$ time, register 0 & TR  \\ \hline
$\xrm_{7}$ & $T_2$ time, register 1& TR   \\ \hline
$\xrm_{8}$ & $T_2$ time, register 2& TR   \\ \hline
$\xrm_{9}$ & $T_2$ time, register 3& TR   \\ \hline
$\xrm_{10}$ & $T_2$ time, register 4& TR   \\ \hline
$\xrm_{11}$ & $H$ fidelity, register 0& CP   \\ \hline
$\xrm_{12}$ & $H$ fidelity, register 1& CP   \\ \hline
$\xrm_{13}$ & $H$ fidelity, register 2& CP \\ \hline
$\xrm_{14}$ & $H$ fidelity, register 3  & CP  \\ \hline
$\xrm_{15}$ & $H$ fidelity, register 4  & CP \\ \hline
\end{tabular}
\medskip\medskip\medskip
\captionsetup{font=small}
\caption{The 16-parameter model derived from the washington data set is composed from noise processes  `ABC': asymmetric binary channel, `CP': coherent phase error model, 'DP': depolarizing noise and `TR': thermal relaxation. The two-qubit model `DP$\otimes$DP' is a tensor product of depolarizing noise.}
\label{tab:noiseParameters}
\end{table}
\subsection{Test circuit}
We validate the stability bound using a numerical simulation of of the Bernstein-Vazirani circuit \cite{bernstein1993quantum} with a noise model using the characterization data presented in the previous section. The Bernstein-Vazirani algorithm determines a secret $n$-bit string $r$ encoded in an oracle. Our focus is on assessing the success probability of correctly computing the secret bit string using the fewest number of queries possible. In contrast to the classical approach that requires $n$ queries, the Bernstein-Vazirani algorithm achieves the same outcome with just one query. Fig.~\ref{fig:bv_qiskit_ckt} illustrates the quantum circuit corresponding to a 4-bit secret key. The observable for the problem is $O~=~\Pi_r~=~\ket{r}\bra{r}$ where $\ket{r} = \bigotimes\limits_{i=1}^n \ket{r_i}$ with $r_i \in \{0,1\}$. 
\subsection{Method}
\begin{figure}[!t]
  \centering
\includegraphics[width=.84\columnwidth]{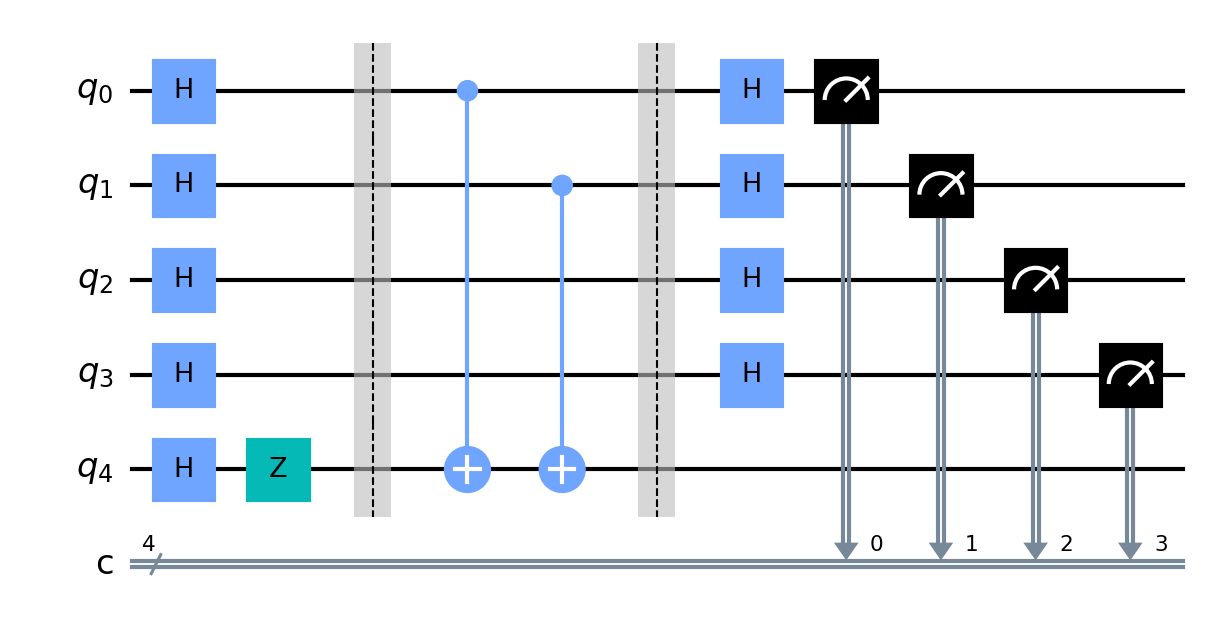}
\caption{The Bernstein-Vazirani circuit for a 4-bit secret string $r = 0011$. $H$ represents the Hadamard gate and $Z$ represents a phase shift of 180 degrees to the $\ket{1}$ state. The meter symbols are measurement operation that project into the computational basis and record results in a classical register $c$.}
\label{fig:bv_qiskit_ckt}
\end{figure}

We used numerical simulations to test the reliability of a model noisy quantum device and to investigate the boundedness of the stability metric as predicted by the theory above. We first mapped the 16 noise parameters necessary for simulating the 5-qubit Bernstein-Vazirani circuit shown in Fig.~\ref{fig:bv_qiskit_ckt} to distinct independent noise processes, selecting them based on the criteria outlined in \cite{divincenzo2000physical} for the physical implementation of quantum computing. 
The parameters mapped to gate and register specific noise model in Table~\ref{tab:noiseParameters}. For example, the asymmetric binary channel for register $0$ flips the measured output bit $b_0$ to $b_0\oplus 1$ with probability $x_0$, while the coherent phase error channel for the Hadamard gate $H$ applied to register 0  transforms the underlying quantum state as $CP(H \rho H) = R_z(\theta) H \rho H R_z^\dagger(\theta)$. 
Thermal relaxation is modeled by an exponential dephasing process that depends on the $T_2$ time and the duration of the underlying gate not shown here.
\par
\begin{figure}[!h]
  \centering
\includegraphics[width=.84\columnwidth]{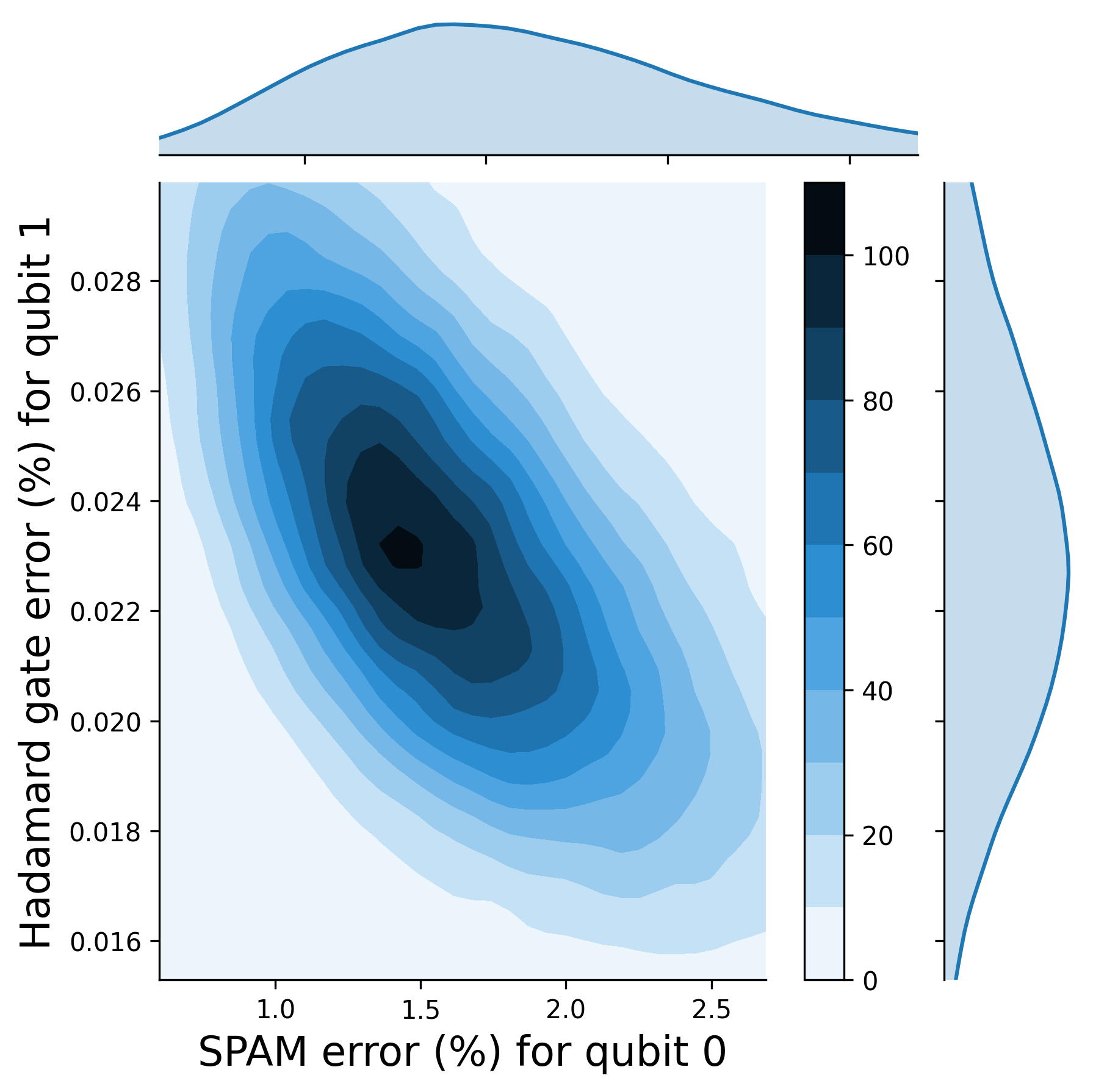}
\caption{A two-dimensional projection of the 16-dimensional correlated distribution modeled using the copula function. These parameter densities present SPAM error on qubit 0 and Hadamard gate error on qubit 1 for Jan-2022.}
\label{fig:dist_with_copula}
\end{figure}

 
While the 16 noise processes above act independently, the underlying noise parameters are assumed to be correlated. We construct a joint distribution of to describe these parameters using the method of Gaussian copula, cf. Fig.~\ref{fig:dist_with_copula}. The copula itself is defined as
\begin{equation}
\Theta(y) = \frac{\exp\left( -\frac{1}{2}(y-\mu)^T \Sigma^{-1} (y-\mu)\right)}{(2\pi)^{n/2}|\Sigma|^{1/2}}
\end{equation}
where the correlation elements $\Sigma_{i,j}$ are derived from Pearson correlation coefficients computed using the daily metric values available from the washington data set. For a Gaussian copula, the corresponding 16-dimensional noise distribution takes the form
\begin{equation}
f_{X}(\xrm;t) = \Theta \left[ F_{X_1}(\xrm_1; t), \cdots F_{X_{d}}(\xrm_{d}; t) \right]
\prod\limits_{j=1}^{n} f_{X_j}(\xrm_j; t)
\label{eq:copulas}
\end{equation}
where $F_{X}(\xrm; t)$ is the cumulative distribution function at time $t$ and $\Theta(\cdot)$ is the copula function. These generated distributions are then used to calculate the Hellinger distance in Eq.~\ref{eq:hellinger_unmodified}. 
\par 
Our numerical studies of the quantum circuit stability generates an ensemble of noisy simulations by drawing samples from the multi-parameter noise distribution represented by Eq.~\ref{eq:copulas}.
We initially establish a joint distribution from the daily data gathered in January 2022 for the washington device, utilizing copulas. Over the next 15 months, we introduce minor perturbations to this distribution, ensuring that the Hellinger distance never exceeds $H_\text{max}$ between the perturbed and original January 2022 distributions. In this perturbation scheme, the marginal distribution of the CNOT gate error between qubits 1 and 2 is modeled using a beta distribution, which is based on the aforementioned January 2022 daily data. Small, random perturbations to the beta distribution parameters are incorporated over 15 months for the CNOT error, with Hellinger distance constraint maintained. For each perturbed distribution, we generate 100,000 noise metric samples, and execute 100 qiskit simulations (each with 8192 shots). The stability metric is then computed from the obtained output, as per Eqn.~\ref{eq:stability}.
%
\subsection{Results}
\begin{figure}[!t]
  \centering
\includegraphics[width=.84\columnwidth]{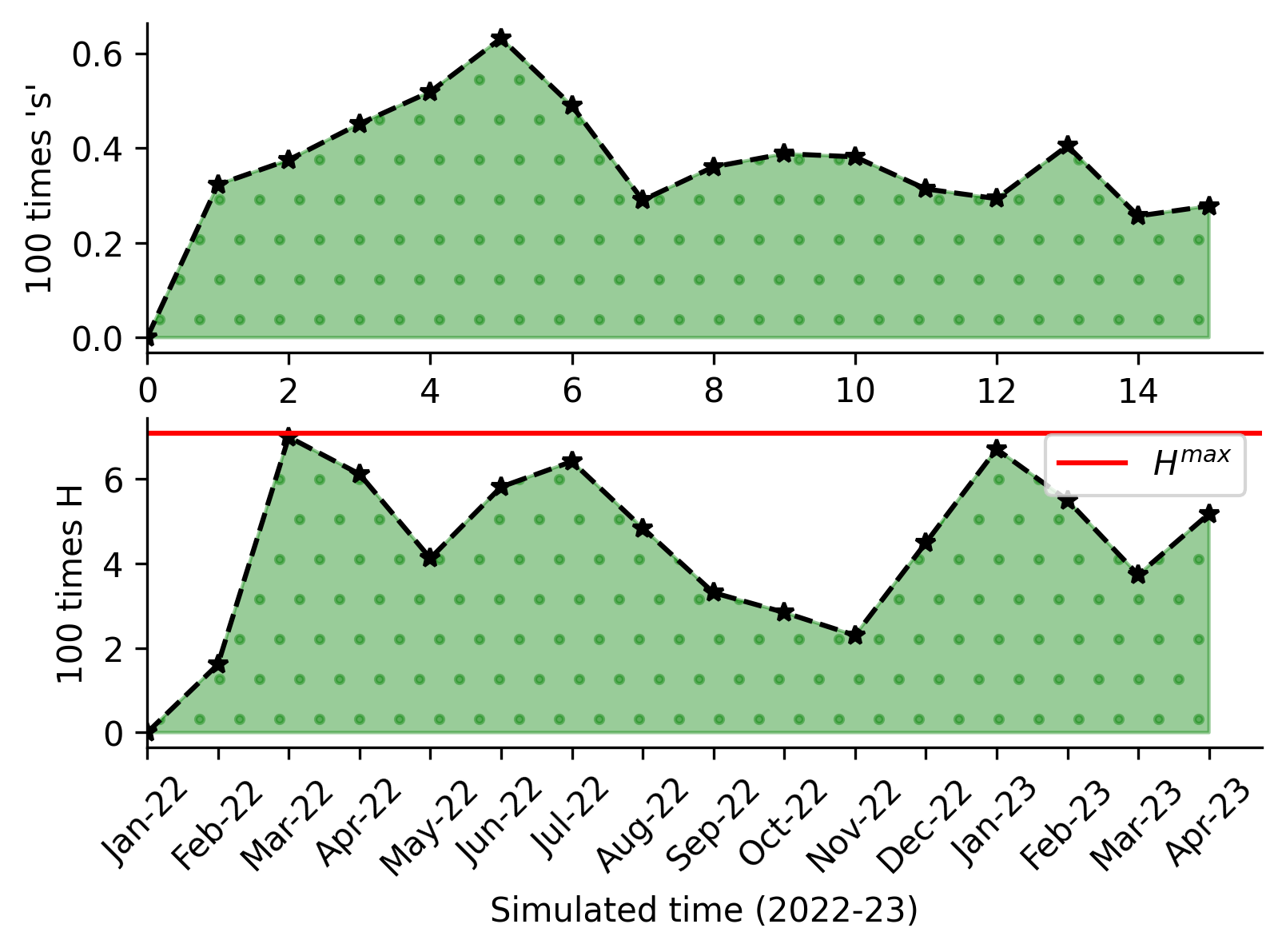}
\caption{Simulation demonstrating that when $H\leq H_\textrm{max}$ (i.e. a reliable, slowly varying noise platform), then $s\leq s_\textrm{max}$ (i.e. the temporal difference of the observable stays within the predicted upper bound).}
\label{fig:slowly-varying}
\end{figure}

Figure~\ref{fig:slowly-varying} presents the simulation results illustrating the relationship between the stability metric ($s$) and the reliability of a quantum device characterized by the Hellinger distance ($H$). The results demonstrate that when $H\leq H_\textrm{max}$ the device is reliable such that the temporal difference of the observable ($s$) remains within the specified upper bound ($s\leq s_\textrm{max}$). 
\par
In our simulations, we set the tolerance threshold $s_\text{tol} = 20\%$, which limits the maximum acceptable deviation in the expectation value over time. According to Eqn.~\ref{eq:bound}, this results in an upper limit of 7.1\% for the device reliability metric $H_\text{max}$. The lower panel of Fig.~\ref{fig:slowly-varying} presents the calculated Hellinger distance between the multi-dimensional noise processes characterizing the device. These calculations show how noise fluctuates on a monthly basis while still respecting the $H_\text{max}$ constraint. While time varying, these process emulate the behavior of a reliable device. The upper panel of Fig.~\ref{fig:slowly-varying} presents the corresponding stability metric, which never exceeds the 20\% tolerance. Moreover, we find the stability is nearly two orders of magnitude smaller than the expected tolerance, with an average of approximately 0.6\%. By selecting a reliable device whose temporal noise variation remains within the defined bounds, we can ensure the stability of program output within the desired tolerance.
\section{Conclusions}
Output stability is crucial in quantum computing research as non-stationary noise processes in quantum devices can result in unstable results that fluctuate based on time-varying device noise characteristics, rendering them unsuitable for meaningful interpretation and drawing scientific conclusions. The variations in superconducting qubits, attributed to certain oxides on the superconductors' surface modeled as fluctuating two-level systems (TLS) \cite{muller2015interacting, klimov2018fluctuations}, have been extensively studied. Ongoing research focuses on addressing the time-varying nature of quantum noise through modeling \cite{etxezarreta2021time}, characterizing noise sources, tracking their temporal profile \cite{proctor2019detecting}, 
and integrating statistical knowledge into quantum architectures using innovative techniques \cite{danageozian2022recovery}. This paper explores the relationship between device reliability and output stability by considering a user-defined upper bound on variations in expectation values. The goal is to assess the stability of program outputs by evaluating the reliability metric within a specified tolerance bound through simulations.
%
\section*{ACKNOWLEDGMENTS}
\small{This work is supported by the U.~S.~Department of Energy (DOE), Office of Science, Early Career Research Program. This research used computing resources of the Oak Ridge Leadership Computing Facility, which is a DOE Office of Science User Facility supported under Contract DE-AC05-00OR22725. }
\bibliographystyle{unsrt}
\bibliography{references.bib}
\end{document}